
\def\bh{black hole }
\def\la{\lambda}
\def\laq{\lambda^2}
\def\section#1{\bigskip\noindent{\bf#1}\smallskip}

\magnification=1200

\font\titolino=cmbx10
\font\tsnorm=cmr10
\font\tscors=cmti10

\font\tscorsp=cmti9
\magnification=1200

\def\PRD{{\tscors Phys. Rev. D }}

\def\NPB{{\tscors Nucl. Phys. B }}

\def\PLB{{\tscors Phys. Lett. B }}

\def\CQG{{\tscors Class. Quantum Grav. }}

\def\note{\advance\notenumber by 1 \footnote{$^{\the\notenumber}$}}
\def\ref#1{\medskip\everypar={\hangindent 2\parindent}#1}
\def\beginref{\begingroup
\bigskip
\leftline{\titolino References.}
\nobreak\noindent}
\def\endref{\par\endgroup}
\def\beginsection #1. #2.
{\bigskip
\leftline{\titolino #1. #2.}
\nobreak\noindent}

\nopagenumbers
\null
\vskip 5truemm
\rightline {INFNCA-TH9618}

\rightline { }
\rightline { }
\rightline{September 1996}
\vskip 15truemm
\centerline{\titolino THE 2D ANALOGUE OF THE REISSNER-NORDSTROM SOLUTION}
\vskip 15truemm
\centerline{\tsnorm S. Monni and  M. Cadoni}
\bigskip
\centerline{\tscorsp Dipartimento di Scienze Fisiche,}
\smallskip
\centerline{\tscorsp Universit\`a  di Cagliari,}
\smallskip
\centerline{\tscorsp Via Ospedale 72, I-09100 Cagliari, Italy.}
\smallskip
\centerline {\tscorsp and}
\smallskip
\centerline{\tscorsp INFN, Sezione di Cagliari.}
\bigskip
\vskip 15truemm
\centerline{\tsnorm ABSTRACT}
\begingroup\tsnorm\noindent
\baselineskip=2\normalbaselineskip
A two-dimensional (2D) dilaton gravity model, whose static solutions have 
the same features of the Reissner-Nordstrom solutions, is obtained from 
the dimensional reduction of a four-dimensional (4D) string effective 
action invariant under 
S-duality transformations. The \bh solutions of the 2D model and their 
relationship with those of the 4D theory are discussed.  


\vfill
\leftline{\tsnorm PACS: 04.70.Bw, 04.60.Kz\hfill}
\smallskip
\hrule
\noindent
\leftline{E-Mail: CADONI@CA.INFN.IT\hfill}
\bigskip
\endgroup
\vfill
\eject
\footline{\hfill\folio\hfill}
\pageno=1

Dilaton gravity in two space-time dimensions  has  been widely 
investigated in recent years [1-7]. The realization that most models of 
two-dimensional (2D) dilaton gravity admit solutions describing black 
holes  has raised the hope to find a mathematically tractable 
setting for  problems of four-dimensional (4D) 
black hole physics that otherwise are very hard to tackle. 
It has been found that 2D dilaton gravity models 
represent in several cases a good and tractable approximation for 
the S-wave sector of 4D black holes [2,4,5]. For example, 
retaining only the 
radial modes of 4D Einstein gravity one gets 2D spherically symmetric 
gravity (see for instance ref. [2]). The celebrated, string inspired, 
Callan-Giddings-Harvey-Strominger (CGHS) model [3] can be also thought of as 
describing  the S-wave sector of 4D dilatonic black holes near 
extremality [4].
In a similar way, it has been shown that also 2D dilaton gravity models 
admitting 
anti-de Sitter black hole solutions can be used to model the physics 
of a class of string-inspired 4D dilatonic black holes [5].

The purpose of this letter is to present the \bh solutions of a 2D 
dilaton gravity model that can be considered as the 2D analogue of the 
Reissner-Nordstrom (RN) solutions of  the 4D Einstein-Maxwell theory.
The 2D model describes the S-wave sector of the 4D dilatonic black holes 
that have been found in ref. [8] in the context of an S-duality model.

In ref. [8], we considered a model described by the action:

$$A=\int d^4x\sqrt{-\hat g}\left[R[\hat g] -2(\hat\nabla\phi)^2-
\cosh(2g\phi)F^2\right].\eqno(1)$$
When the field $\phi$ is referred to as the dilaton, the action 
turns out to be invariant under an
S-duality transformation and 
 can be considered as a
generalization of the usual  functional
$$A= \int d^4x\sqrt{-\hat g}\left[R[\hat g]-2(\hat \nabla\phi)^2-e^{- 2\phi}F^2\right],$$
which is part of the action describing the low-energy dynamics of string theory.
Moreover, letting $\phi$ be a modulus field, the action (1) can be thought of
 as an approximation
to the effective action resulting from toroidal or orbifold compactification. 
We were 
able to find exact, magnetically charged,  black hole solutions of the model 
for $g^2=1$ and $g^2=3$. In particular, the $g^2=1$ solution
can be put into the form:
 $$e^{2\phi}= e^{2\phi_0}(1+{2\sigma\over r}),\eqno (2a) $$
 $$F=Q\sin{\theta}{d\theta}\wedge {d\varphi},\eqno(2b) $$
 $$ds^2=-{(r-r_-)(r-r_+)\over r(r+2\sigma)}\, dt^2+
{r(r+2\sigma)\over (r-r_-)(r-r_+)}\,dr^2 + r(r+2\sigma) d 
\Omega^2,\eqno(2c) $$
where $r_{\pm}=M-\sigma \pm 
\sqrt{M^{2}+\sigma^{2}-Q^{2}\cosh{(2\phi_{0})}}$
and the integration constants have to satisfy the equation $2M\sigma=-Q^{2} 
\sinh (2\phi_{0})$.
This solution shares its main features with the RN solution 
of general relativity. Moreover, it reduces to the 
Garfinkle-Horowitz-Strominger  (GHS) solution [9] 
in the $\phi_0 \to -\infty$ weak coupling regime [8] (due to the 
duality symmetry of the action  also in the strong coupling regime) . 

Let us write the action (1) in terms of the usual "string" metric 
$g=\exp(2\phi)\hat g$:
$$A=\int d^4x\sqrt{-g}\left[\left(R+4(\nabla\phi)^2\right)e^{-2\phi}
-\cosh(2g\phi)F^2\right].\eqno(3)$$
For ${g^2}=1$, the action (3) admits   magnetically 
charged black hole solutions 
whose topology is $G^2\times S^2, $ 
where $G^2$ is an asymptotically flat 2D space-time and the radius of the 
two-sphere $S^{2}$ is  constant and equal to the magnetic charge. 
In fact,  the field equations derived from the action  (3), with $g^2=1$,
have solutions of the form:
$$ ds^2=ds_{(2)}^2+{1\over \laq }d\Omega^2,\eqno(4) $$
with $\la=1/Q$, the Maxwell tensor given as in eq. (2b) and the dilaton
depending only on the coordinates of $G^{2}$.
From the 4D field equations follows that the metric over $ G^2$ 
and the dilaton $\phi$ must be
solutions of the following equations:
$$\eqalign{R 
+4{\nabla}^2\phi-4(\nabla\phi)^2+\laq\left(1+e^{4\phi}\right) =0,
\cr \nabla_{\mu}\nabla_{\nu}\phi -g_{\mu\nu}\left( \nabla^2\phi-(\nabla\phi)^2
\right)-{1\over 4}
\laq g_{\mu\nu}(1-e^{4\phi})&=0,\cr}\eqno (5)$$
where all the objects are calculated with respect to the 2D metric over
$G^{2}$.
The field equations (5) can be derived from the  dimensionally reduced 
action-functional:
$$A={1\over 2 \pi}\int d^2x\sqrt{-g} \left [e^{- 2\phi}
\left(R+4(\nabla\phi)^2
\right) -2\laq\sinh 2\phi\right].\eqno(6)$$
This 2D action can be obtained  by substituting the ansatz (2b), (4) 
directly into the 4D action (3).

The 2D black hole solutions of the field equations (5) can be easily 
obtained; they have the following form :

$$\d s^2=-(1-a_{+}e^{-\la x})(1-a_{-}e^{-\la x})\d t^2+
(1-a_{+}e^{-\la x}){^{-1}}(1-a_{-}e^{-\la x}){^{-1}}\d x^2, \eqno(7)$$
$$\phi=\phi_0 -{\la\over{2}}x .\eqno(8) $$
The parameters $a_{\pm}$ and $\phi_0$ appearing in the previous 
equations are not independent but
constrained by the equation:
$$a_{+}a_{-}=e^{4\phi_{0}}.\eqno(9)$$
Apart from the constant $\phi_{0}$, the solution is parameterized by the value of 
a single physical observable, which can be identified with the mass 
of the black hole. This result is a consequence of Birkhoff's theorem for 2D 
dilaton gravity [6]. The mass $m$ of the solution (7)
 can be calculated using the 
procedure due to Mann [7]; one finds
$$m={\la\over 2}(a_{+}+a_{-})\eqno(10).$$
Using eqs. (9) and (10), one can express the parameters $a_{\pm}$  as a 
function of the physical observables: 
$$a_{\pm}={1\over 
\la}\left(m\pm\sqrt{m^{2}-\laq e^{4\phi_{0}}}\right).\eqno(11)$$
The expressions (9), (10) and (11) are 
similar to those one obtains in the RN case, with $\la \exp(2\phi_0)$ playing 
the role of the 4D magnetic charge.
For $m>\la e^{2\phi_{0}},$ eq. (7)  describes  an asymptotically flat 
2D  space-time with an 
inner and 
outer horizon at $x=x_{\pm}=(1/\la)\ln a_{\pm},$
respectively. For $m=\la e^{2\phi_{0}}$ the black hole becomes extremal. The point $x=-\infty$ is a curvature singularity 
analogous to the origin of the radial coordinate system in the RN 
case. The space-time has the same causal structure as the RN solution 
both in the general and in the  extremal case. This can be easily 
demonstrated by noting that for $m>\la e^{2\phi_{0}}$ the coordinate 
transformation
$${|e^{\la x}-a_{+}|^{a_{+}}\over |e^{\la x}-a_{-}|^{a_{-}}}=
e^{\la(a_{+}-a_{-})y},$$
brings the metric (7) into the form
$$ds^{2}=\left(1-a_{+}e^{-\la x }\right) \left(1-a_{-}e^{-\la x }\right)
\left(-dt^{2}+dy^{2}\right).$$
In the extremal case $a_{+}=a_{-}=a$, the coordinate transformation 
involved is 
$$\la y=\ln | e^{\la x}- a| - {a\over e^{\la x}-a}\,.$$

The 2D black hole solution (7) describes the S-wave sector of the 
near-extreme 4D black hole solution of the theory defined by the 
action (3). The latter solution can be easily obtained from the solution 
(2) and the Weyl rescaling leading to the action (3). 
Introducing a new radial coordinate $\rho$,
$$\Delta \sinh^{2} \rho= r-r_{+}, \qquad \Delta=r_{+}-r_{-},$$
one finds that in the extremal limit $\Delta\rightarrow 0$, $\Delta 
e^{-\phi}=const,$ the Weyl-rescaled solution behaves as follows:
$$ds^{2}=- \Delta^{2}{\cosh^{2}\rho \,\sinh^{2}\rho\over 
(r_{+}+\Delta 
\sinh^{2}\rho)^{2}}\,dt^{2}+Q^{2}(4d\rho^{2}+d\Omega^{2}),\eqno(12a)$$
$$e^{2\phi}={Qe^{\phi_{0}}\over r_{+}+\Delta 
\sinh^{2}\rho}\,.\eqno(12b)$$
Performing the  coordinate transformation 
$\Delta \sinh^{2}\rho= Q\exp(-\phi_{0}) \exp(\la x) -r_{+},$ after 
the identification $r_{\pm}= Q\exp{(-\phi_{0})} a_{\pm}$, one can 
easily realize that the solution (7), (8) coincides with the 2D section 
of the solution (12).
  
It is important to notice that an extremal limit in which the 4D solution 
behaves like eqs. (12)  exists only 
for $\sigma>0$, $\phi_{0}<0$, i.e in the weak coupling region of the 
theory. For  $\sigma<0$, $\phi_{0}>0$, i.e in the strong coupling 
region, there is no limit in which the topology of the  4D solutions 
factorize as $G^{2}\times S^{2}$. The S-duality symmetry 
$\phi\rightarrow - \phi$ of the action (1) is not preserved by the 
Weyl rescaling leading to the action (3). In terms of the "string" 
metric, the strong ($\phi_{0}\rightarrow \infty$)  and weak ($\phi_{0}
\rightarrow -\infty)$ coupling regimes admit a different description. 
Only for the latter an approximate solution like  (12) exists.
Similarly, the 2D theory described by the action (6) is not invariant 
under the duality transformation $\phi\rightarrow - \phi$. 
In the weak coupling regime the action (6) describes the CGHS model.
This is what one would have expected because  the CGHS model describes
the S-wave sector of the GHS black hole near extremality [4] and 
because the solutions (2) 
reduce in the weak coupling regime to the GHS solutions.
From eqs. (7) and (11) it is evident that in the weak coupling regime 
$\phi_{0}\rightarrow 
-\infty$ the inner horizon disappears and that the solution becomes 
the CGHS solution:
$$ds^{2}=-\left(1-{2M\over \la}e^{-\la x}\right)dt^{2}+\left(1-
{2M\over \la}e^{-\la x}\right)^{-1}dx^{2},$$
$$\phi=\phi_{0}-{\la\over 2} x.$$

We conclude this letter with some remarks on the existence of 
magnetically charged solutions with topology $G^{2}\times S^{2}$ 
in the context of 4D 
dilaton gravity theories. Solutions of this kind, apart from the case 
we have discussed in this paper, have been  already found in several 
cases: the Einstein-Maxwell theory 
(the Bertotti-Robinson solutions [10]),
4D  string effective theory [4,9], the string-inspired models of 
ref. [5,11]. One could be led to think that the existence of this kind of 
solutions is a general feature of 4D dilaton gravity theories. However, 
it turns out that the opposite of this statement is true: the 
existence of magnetically charged solutions with topology 
$G^{2}\times S^{2}$ implies a strong constraint on the form of the 
gauge coupling function between the Maxwell field and the dilaton.
Let us consider the general action
$$A_{4}=\int d^4x\sqrt{-g}\bigl[D(\phi)R+H(\phi)(\nabla\phi)^2+
V(\phi)F^2\bigr], \eqno(13)$$
where $D(\phi)$, $H(\phi)$ and $V(\phi)$ are coupling functions.
Its extremals are  solutions of the following equations:
$$\nabla_\nu[V F^{\mu\nu}]=0,\eqno(14)$$ 
$${dD\over d\phi}R-{dH\over d\phi}(\nabla{\phi})^{2}+
{dV\over d\phi}F^{2}-2H\nabla^{2}\phi=0,\eqno(15)$$
$$DR_{\mu\nu}+H\nabla_{\mu}\phi\nabla_{\nu}\phi -\nabla_{\mu}
\nabla_{\nu}D-{1\over2}g_{\mu\nu}\nabla^{2}D+
2V\left(F_{\mu\alpha}F_{\nu}^{\alpha}-{1\over4}g_{\mu\nu}F^{2}\right) 
=0.\eqno(16)$$
If one substitutes the ansatz (2b) and (4) into these equations and considers 
$\phi$ to be a function only of the coordinates of $ G^2$, 
eq. (15) and the first two equations in (16) 
reduce to the equations of motion
coming from  the dimensionally reduced action:
$$A_{2}={1\over 2\pi}\int d^2x\sqrt{-g}\left[D(\phi)R+H(\phi)(\nabla\phi)^2+
2\la^{2}D(\phi)+2\la^{4}Q^{2}V(\phi)\right], $$  
obtained by inserting
the ansatz directly into $A_{4}.$ The last equation in (16) states that 
the spherical component
of the energy tensor of the 4D theory
is a first integral. This equation together with eq. (15) 
represents a strong constraint on the form of the coupling 
function $V(\phi)$.
Taking as usual $D=\exp(-2\phi)$ and $H=4\exp(-2\phi)$, 
the general solution  of the constraint can be parameterized 
in terms of two arbitrary constants $A, B$ 
 and reads:
$$V=A\exp(2\phi)+B\exp(-2\phi),$$
with $B<0.$  
Thus even for an action of the form (3) magnetically charged solutions 
with topology $G^{2}\times S^{2}$ are allowed only for the value of the 
coupling constant $g$ we considered here, namely $g^{2}=1$.

\bigskip

\beginref           
\ref [1]  T. Banks and M. O'Loughlin, \NPB {\bf 362}, 649 (1991) and
 \PRD {\bf 48}, 698 (1993); 
D.A. Lowe, M. O'Loughlin, \PRD {\bf 48}, 3735 (1993);
S.P. Trivedi, \PRD {\bf 47}, 4233 (1993); 
K.C.K. Chan and R.B. Mann, \CQG {\bf 12}, 1609 (1995); M. Cadoni, S. 
Mignemi \PLB {\bf 358}, 217 (1995);  
M. Cadoni, \PRD {\bf 53}, 4413 (1996).

\ref [2] A. Strominger in {\sl Les Houches Lectures on
Black holes}, lectures presented at the 1994 Les Houches Summer 
School, HEP-TH/9501071.

\ref [3] C.G. Callan, S.B. Giddings, J.A. Harvey and A. Strominger, 
\PRD {\bf 45}, 1005 (1992).
    
\ref [4] S.B. Giddings and A. Strominger, 
\PRD {\bf 46}, 627 (1992).

\ref [5] M. Cadoni and S. Mignemi, \NPB {\bf 427}, 669 (1994) and
 \PRD {\bf 51}, 4319 (1995).

\ref [6] D. Louis Martinez and G. Kunstatter, \PRD {\bf 49}, 5227 (1994).

\ref [7] R.B. Mann, \PRD {\bf 47}, 4438  (1993).

\ref [8] S. Monni and M. Cadoni, \NPB {\bf 466}, 
 101 (1996).

\ref [9] D. Garfinkle, G. T. Horowitz and  A. Strominger, 
\PRD {\bf 45}, 3140 (1991).

\ref [10] B. Bertotti, {\tscors Phys. Rev.} {\bf 116}, 1331 (1959);
I. Robinson, {\tscors Bull. Acad. Polon. Sci.} {\bf 7}, 351 (1959).

\ref [11] M. Cadoni and S. Mignemi, \PRD {\bf 48}, 5536 (1993).

\endref

\end